\documentclass[aps,prc,twocolumn,floatfix,nofootinbib,showpacs,superscriptaddress]{revtex4-1}

\usepackage{amsmath,amsfonts,amssymb,bm}
\usepackage{graphicx}
\usepackage{color}
\usepackage{pzccal}
\usepackage{soul}
\definecolor{purple}{rgb}{0.5,0,0.5}
\definecolor{blue}{rgb}{0.0,0,0.9}

\begin{document}

\title{Imaging dynamical chiral symmetry breaking: pion wave function on the light front
}

\author{Lei Chang}
\affiliation{Institut f\"ur Kernphysik, Forschungszentrum J\"ulich, D-52425 J\"ulich, Germany}

\author{I.\,C.~Clo\"et}
\affiliation{CSSM and CoEPP, School of Chemistry and Physics
University of Adelaide, Adelaide SA 5005, Australia}
\affiliation{Physics Division, Argonne National Laboratory, Argonne, Illinois 60439, USA}

\author{J.\ J.~Cobos-Martinez}
\affiliation{Center for Nuclear Research, Department of
Physics, Kent State University, Kent OH 44242, USA}
\affiliation{
Departamento de F\'{i}sica, Universidad de Sonora, Boulevard Luis Encinas J. y Rosales, Colonia Centro, Hermosillo, Sonora 83000, Mexico}

\author{\mbox{C.\,D.~Roberts}}
\affiliation{Physics Division, Argonne National Laboratory, Argonne, Illinois 60439, USA}
\affiliation{Department of Physics, Illinois Institute of Technology, Chicago, Illinois 60616-3793, USA}

\author{S.\,M.~Schmidt}
\affiliation{Institute for Advanced Simulation, Forschungszentrum J\"ulich and JARA, D-52425 J\"ulich, Germany}

\author{P.\,C.~Tandy} \affiliation{Center for Nuclear Research, Department of
Physics, Kent State University, Kent OH 44242, USA}


\date{2 January 2013}

\begin{abstract}
%
We project onto the light-front the pion's Poincar\'e-covariant Bethe-Salpeter wave-function, obtained using two different approximations to the kernels of QCD's Dyson-Schwinger equations.  At an hadronic scale both computed results are concave and significantly broader than the asymptotic distribution amplitude, $\varphi_\pi^{\rm asy}(x)=6 x(1-x)$; e.g., the integral of $\varphi_\pi(x)/\varphi_\pi^{\rm asy}(x)$ is $1.8$ using the simplest kernel and $1.5$ with the more sophisticated kernel.  Independent of the kernels, the emergent phenomenon of dynamical chiral symmetry breaking is responsible for hardening the amplitude.
\end{abstract}

\pacs{
12.38.Aw,   
12.38.Lg,   
11.10.St,	
14.40.Be	
}

\maketitle

%
%
The momentum-space wave-function for a nonrelativistic quantum mechanical system, $\psi(p,t)$, is a probability amplitude, such that $|\psi(p,t)|^2$ is a non-negative density which describes the probability that the system is described by momenta $p$ at a given equal-time instant $t$.  Although the replacement of certainty in classical mechanics by probability in quantum mechanics was disturbing for some, the step to relativistic quantum field theory is still more confounding.  Much of the additional difficulty owes to the loss of particle number conservation when this step is made: two systems with equal energies need not have the same particle content, because that is not conserved by Lorentz boosts, so that even interpretation via probability densities is typically lost.  To exemplify: a charge radius cannot generally be defined via the overlap of two wave-functions because the initial and final states do not possess the same four-momentum and hence are not described by the same wave-function.

Such difficulties may be circumvented by formulating a theory on the light-front because the eigen-functions of the light-front Hamiltonian are independent of the system's four-momentum \cite{Keister:1991sb,Brodsky:1997de}.  The light-front wave-function of an interacting quantum system therefore provides a connection between dynamical properties of the underlying relativistic quantum field theory and notions familiar from nonrelativistic quantum mechanics.  It can translate features that arise purely through the infinitely-many-body nature of relativistic quantum field theory into images whose interpretation is seemingly more straightforward.  Naturally, that is only achieved if the light-front wave-function can be calculated.

A phenomenon for which a quantum mechanical image would be desirable is dynamical chiral symmetry breaking (DCSB).  Strictly impossible in quantum mechanics with a finite number of degrees-of-freedom, this striking emergent feature of quantum chromodynamics (QCD), the strong-interaction part of the Standard Model, plays a critical role in forming the bulk of the visible matter in the Universe \cite{national2012Nuclear}.  Expressed in numerous aspects of the spectrum and interactions of hadrons; e.g., the large splitting between parity partners \cite{Chang:2011ei,Chen:2012qrS} and the existence and location of a zero in some hadron form factors \cite{Wilson:2011aa}, DCSB has not yet been realised in the light-front formulation of quantum field theory.

The impact of DCSB is expressed with particular force in properties of the pion.  It is the pseudo-Goldstone boson that emerges when chiral symmetry is dynamically broken, so that its very existence as the lightest hadron is grounded in DCSB.  As a corollary, numerous model-independent statements may be made about the pion's Bethe-Salpeter amplitude and its relationship to the dressed-quark propagator \cite{Maris:1997hd}.  Given that the pion's light-front valence-quark distribution amplitude (PDA) can be computed from these two quantities, their calculation provides a means by which to expose DCSB in a wave-function with quantum mechanical characteristics.

Consider, therefore, the following projection of the pion's Bethe-Salpeter wave function onto the light-front
\begin{equation}
f_\pi\, \varphi_\pi(x) = {\rm tr}_{\rm CD}
Z_2 \! \int_{dq}^\Lambda \!\!
\delta(n\cdot q_+ - x \,n\cdot P) \,\gamma_5\gamma\cdot n\, \chi_\pi(q;P)\,,
\label{pionPDA}
\end{equation}
where: $f_\pi$ is the pion's leptonic decay constant; the trace is over colour and spinor indices; $\int_{dq}^\Lambda$ is a Poincar\'e-invariant regularization of the four-dimensional integral, with $\Lambda$ the ultraviolet regularization mass-scale; $Z_{2}(\zeta,\Lambda)$ is the quark wave-function renormalisation constant, with $\zeta$ the renormalisation scale; $n$ is a light-like four-vector, $n^2=0$; $P$ is the pion's four-momentum, $P^2=-m_\pi^2$ and $n\cdot P = -m_\pi$, with $m_\pi$ being the pion's mass; %
and the pion's Bethe-Salpeter wave-function
\begin{equation}
\chi_\pi(q;P) = S(q_+) \Gamma_\pi(q;P) S(q_-)\,,
\label{chipi}
\end{equation}
with $\Gamma_\pi$ the Bethe-Salpeter amplitude,  $S$ the dressed light-quark propagator, and $q_+ = q + \eta P$, $q_- = q - (1-\eta) P$, $\eta\in [0,1]$.
Owing to Poincar\'e covariance, no observable can legitimately depend on $\eta$; i.e., the definition of the relative momentum.
%
Using Eq.\,\eqref{pionPDA}, one may show that the moments of the distribution; viz., $\langle x^m\rangle := \int_0^1 dx \, x^m \varphi_\pi(x)$, are given by
\begin{equation}
f_\pi (n\cdot P)^{m+1} \langle x^m\rangle =
{\rm tr}_{\rm CD}
Z_2 \! \int_{dq}^\Lambda \!\!
(n\cdot q_+)^m \,\gamma_5\gamma\cdot n\, \chi_\pi(q;P)\,.
\label{phimom}
\end{equation}

The dressed-quark propagator may be expressed
\begin{subequations}
\begin{eqnarray}
S(p)
& = & -i \gamma\cdot p \, \sigma_V(p^2,\zeta^2)+\sigma_S(p^2,\zeta^2)\,, \label{SgeneralSigma}\\
&=& 1/[i \gamma\cdot p \, A(p^2,\zeta^2) + B(p^2,\zeta^2)]\,,
\label{SgeneralNAB}
\end{eqnarray}
\label{SgeneralN}
\end{subequations}
\hspace*{-0.5\parindent}and can be obtained from the gap equation \cite{Maris:2003vk,Bashir:2012fs}:
\begin{eqnarray}
\lefteqn{
\nonumber S^{-1}(p) = Z_2 \,(i\gamma\cdot p + m^{\rm bm})}\\
&& + Z_1 \int^\Lambda_{dq}\!\! g^2 D_{\mu\nu}(p-q)\frac{\lambda^a}{2}\gamma_\mu S(q) \frac{\lambda^a}{2}\Gamma_\nu(q,p) ,
\label{gendseN}
\end{eqnarray}
where: $D_{\mu\nu}$ is the gluon propagator; $\Gamma_\nu$ the quark-gluon vertex; $m^{\rm bm}(\Lambda)$ the current-quark bare mass; and $Z_{1}(\zeta,\Lambda)$ the vertex renormalisation constant.  

The pion's amplitude may be obtained from the Bethe-Salpeter equation, a modern expression of which is explained in Ref.\,\cite{Chang:2009zb}.  With $\eta=1/2$ in the Bethe-Salpeter equation, it is convenient to write the amplitude in the form
\begin{eqnarray}
\nonumber
\lefteqn{\Gamma_{\pi}(q;P) = \gamma_5
\left[ i E_{\pi}(q;P) + \gamma\cdot P F_{\pi}(q;P)  \right. }\\
&&  \left. + \, q\cdot P \gamma\cdot q \, G_{\pi}(q;P) + \sigma_{\mu\nu} q_\mu P_\nu H_{\pi}(q;P) \right], 
\label{genGpi}
\end{eqnarray}
where the functions are even.  Owing to DCSB and the axial-vector Ward-Takahashi identity, all scalar functions in Eq.\,\eqref{genGpi} are nonzero \cite{Maris:1997hd}.  Moreover, in the chiral limit, which we subsequently employ exclusively, $m_\pi=0$ and
\begin{equation}
\label{gtE}
f_\pi E_\pi(q;0) = B(q^2).
\end{equation}
This Goldberger-Treiman-like identity, part of a near complete equivalence between the one-body and pseudoscalar two-body problem in QCD, is a pointwise statement of Goldstone's theorem.  The gap and Bethe-Salpeter equations are key members of the set of Dyson-Schwinger equations (DSEs), which provide an efficacious tool for the study of hadron properties \cite{Maris:2003vk,Bashir:2012fs}.

Significant features of $\varphi_\pi(x)$ in Eq.\,\eqref{pionPDA} can be elucidated algebraically with a simple model before employing numerical solutions for $S(p)$, $\Gamma_\pi$.  To this end, with $\Delta_M(s) = 1/[s+M^2]$ and $\eta = 0$ in Eqs.\,\eqref{pionPDA}, \eqref{chipi}, consider
\begin{eqnarray}
\label{pointS}
S(p) &=& [-i\gamma \cdot p + M] \Delta_M(p^2)\,, \\
\label{rhoznu}
\rho_\nu(z) &=& \frac{1}{\surd \pi}\frac{\Gamma(\nu + 3/2)}{\Gamma(\nu+1)}\,(1-z^2)^\nu\,,\\
\label{rhoEpi}
\Gamma_\pi(q;P) & = &
i\gamma_5 \frac{M^3}{f_\pi} \!\! \int_{-1}^{1}\!\! \!dz \,\rho(z)
\Delta_M^\nu(q_{+z}^2)\,,
\end{eqnarray}
where $q_{\pm z} = q-(1 \mp z )P/2$.
Inserting Eqs.\,\eqref{pointS}--\eqref{rhoEpi} in Eq.\,\eqref{phimom},
using a Feynman parametrisation to combine denominators,
shifting the integration variable to isolate the integrations over Feynman parameters from that over the four-momentum $q$,
and recognising that $d^4q$-integral as the expression for $f_\pi$, one obtains
\begin{equation}
\label{momnum}
\langle x^m \rangle_\nu =
\frac{\Gamma (2 \nu +2) \Gamma (m+\nu +1)}{\Gamma (\nu +1) \Gamma (m+2 \nu +2)}\,.
\end{equation}

Suppose that $\nu =0$; i.e., 
the pion's Bethe-Salpeter amplitude is independent of momentum and hence describes a point-particle, then Eq.\,\eqref{momnum} yields
\begin{equation}
\label{mom0m}
\langle x^m \rangle_0 =
\frac{\Gamma (2) \Gamma (m+1)}{\Gamma (1) \Gamma (m+2)} = \frac{1}{m+1}\,.
\end{equation}
These are the moments of the distribution amplitude
\begin{equation}
\varphi_\pi(x) = 1\,,
\end{equation}
which is indeed that of a pointlike pion \cite{Roberts:2010rnS}.

Alternatively, consider $\nu=1$.  Then $\Gamma_\pi(k^2) \sim 1/k^2$ for large relative momentum.  This is the behaviour in QCD at $k^2\gg \mu_G^2$, where $\mu_G \simeq 0.5\,$GeV is the dynamically generated gluon mass \cite{Ayala:2012pbS}.  $\nu=1$ in Eq.\,\eqref{momnum} yields
\begin{equation}
\label{mom1m}
\langle x^m \rangle_1 = 
\frac{\Gamma (4) \Gamma (m+2)}{\Gamma (2) \Gamma (m+4)} = \frac{6}{(m+3)(m+2)}\,.
\end{equation}
These are the moments of
\begin{equation}
\varphi_\pi^{\rm asy}(x) = 6 x (1-x)\,;
\label{PDAasymp}
\end{equation}
viz., QCD's asymptotic PDA \cite{Lepage:1980fj}.

It is readily established that with Eqs.\,\eqref{pointS}--\eqref{rhoEpi} in  Eq.\,\eqref{phimom} one obtains the ``asymptotic'' distribution associated with a $(1/k^2)^\nu$ vector-exchange interaction; viz.,
\begin{equation}
\label{asymptdistn}
\varphi_\pi(x) = \frac{\Gamma(2\nu + 2)}{\Gamma(\nu+1)^2}\, x^\nu (1-x)^\nu\,.
\end{equation}
Notably, the $z$-modulated dependence on $q\cdot P$ in Eq.\,\eqref{rhoEpi}
is the critical factor in obtaining the results described here.  To illustrate, if one uses $\nu=1$ but $2\rho(z) = \delta(1-z)+\delta(1+z)$, then point-particle moments, Eq.\,\eqref{mom0m}, are obtained even though $\Gamma_\pi(k^2) \sim 1/k^2$ for $k^2 \gg M^2$.  There is a natural explanation.  Namely, with such a form for $\rho(z)$ one assigns equal probability to two distinct configurations: valence-quark with all the pion's momentum and valence-antiquark with none or antiquark with all the momentum and quark with none.  In assigning equal weight to these two extreme configurations one has defined a bound-state with point-particle-like characteristics.  It follows that deviations from the asymptotic distribution may be expressed through $\rho_\nu(z)$.

We solve the gap and pion Bethe-Salpeter equations numerically using the interaction in Ref.\,\cite{Qin:2011ddS}, which preserves the one-loop renormalisation group behaviour of QCD and guarantees that the quark mass-function, $M(p^2)= B(p^2,\zeta^2)/A(p^2,\zeta^2)$, is independent of the renormalisation point, which we choose to be $\zeta=2\,$GeV.  In completing the gap and Bethe-Salpeter kernels we employ two different procedures and compare their results: rainbow-ladder truncation (RL), the most widely used DSE computational scheme in hadron physics, detailed in App.\,A.1 of Ref.\,\cite{Chang:2012cc}; and the DCSB-improved kernels detailed in App.\,A.2 of Ref.\,\cite{Chang:2012cc} (DB), which are the most refined kernels currently available.  Both schemes are symmetry-preserving, and hence ensure Eq.\,\eqref{gtE}, but the latter incorporates essentially nonperturbative effects associated with DCSB into the kernels, which are omitted in rainbow-ladder truncation and any stepwise improvement thereof \cite{Chang:2009zb}.  This kernel thereby exposes a key role played by the dressed-quark anomalous chromomagnetic moment in determining observable quantities \cite{Chang:2010hb} and, e.g., clarifies a causal connection between DCSB and the splitting between vector and axial-vector mesons \cite{Chang:2011ei}.

\begin{figure}[t]
\includegraphics[width=0.80\linewidth]{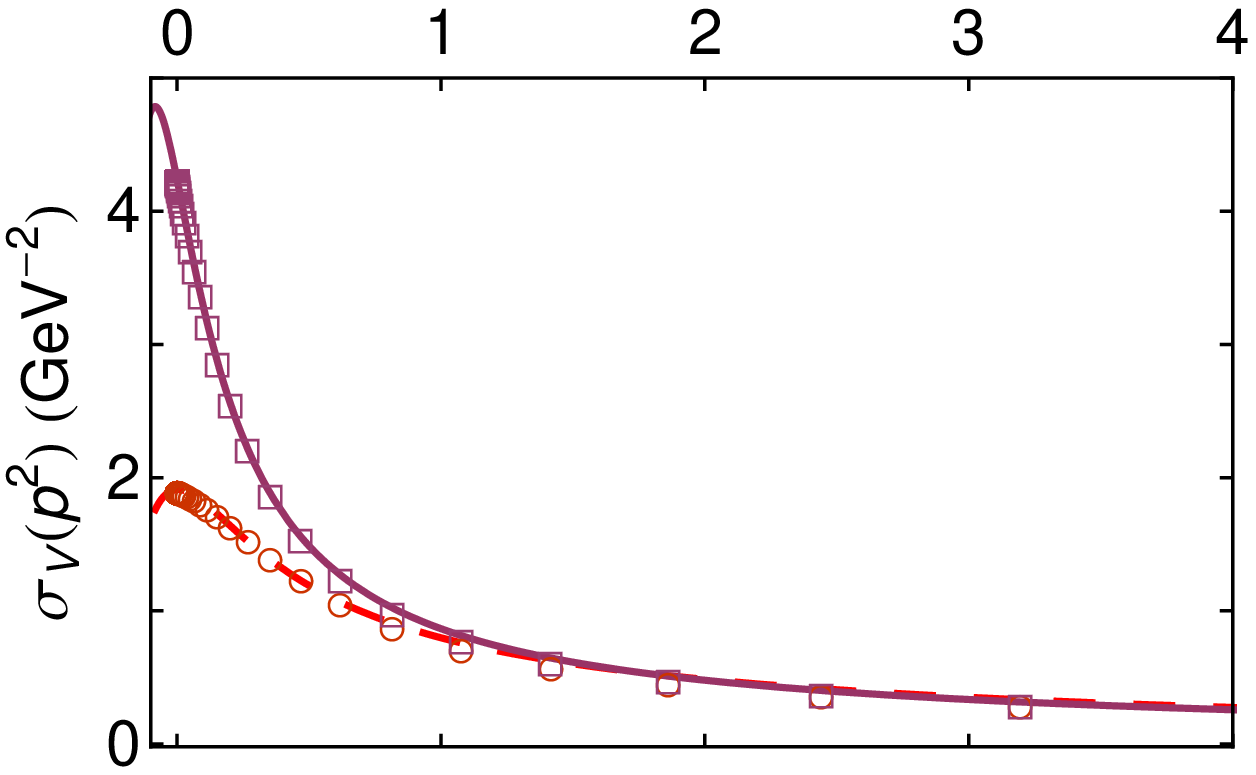}

\vspace*{-8.7ex}
\includegraphics[width=0.8\linewidth]{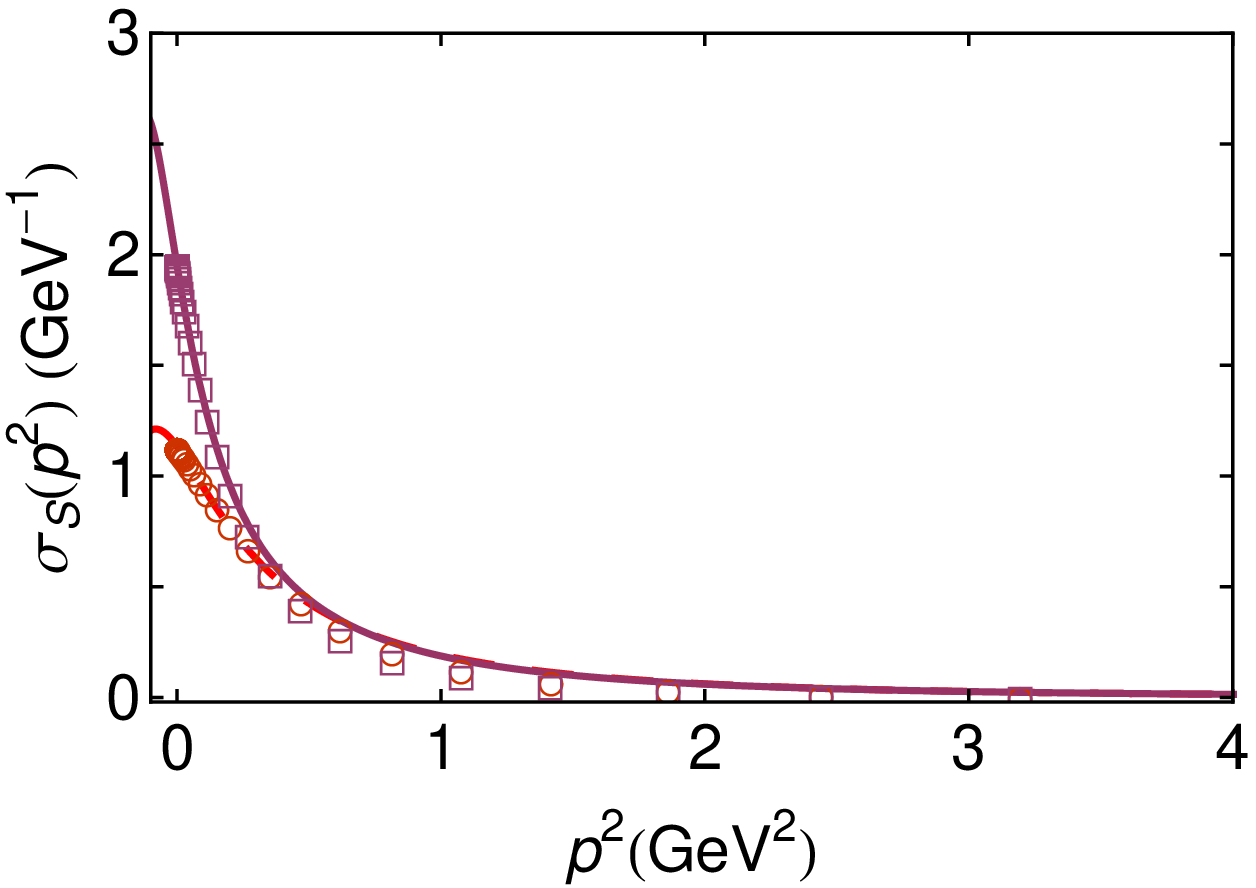}

\caption{Functions characterising the dressed quark propagator.
\emph{Upper panel}. $\sigma_V(p^2)$ -- RL kernel: solution (open circles) and interpolation function (long-dashed curve); and DB kernel: solution (open squares) and interpolation function (solid curve).
\emph{Lower panel}. $\sigma_S(p^2)$, with same legend.
In the chiral limit at large $p^2$, $\sigma_V(p^2)\sim 1/p^2$ and $\sigma_S(p^2) \sim 1/p^4$.
\label{fig:Splot}}
\end{figure}

The solutions are obtained as matrices.  Computation of the moments in Eq.\,\eqref{phimom} is cumbersome with such input, so we employ algebraic parametrisations of each array to serve as interpolations in evaluating the moments.  For the quark propagator, we represent $\sigma_{V,S}$ as meromorphic functions with no poles on the real $p^2$-axis \cite{Bhagwat:2002tx}, a feature consistent with confinement \cite{Bashir:2012fs}.  Each scalar function in the pion's Bethe-Salpeter amplitude is expressed via a Nakanishi-like representation \cite{Nakanishi:1963zz}; i.e., through integrals like Eq.\,\eqref{rhoEpi}, with parameters fitted to that function's first two $q\cdot P$ Chebyshev moments.  (Details are presented in the Appendix.)  The quality of the description is illustrated via the dressed-quark propagator in Fig.\,\ref{fig:Splot}.

Using Eq.\,\eqref{phimom} it is now straightforward to compute arbitrarily many moments of the pion's PDA, $\{\langle x^m\rangle|m=1,\ldots,m_{\rm max}\}$: we typically employ $m_{\rm max}=50$.
Since Gegenbauer polynomials of order $\alpha$, $\{C_n^{\alpha}(2 x -1)| n=0,\ldots,\infty\}$, are a complete orthonormal set on $x\in[0,1]$ with respect to the measure $[x (1-x)]^{\alpha_-}$, $\alpha_-=\alpha-1/2$, they enable reconstruction of any function that vanishes at $x=0,1$.
(N.B.\,
$\varphi_\pi(x)$ is even under $x\leftrightarrow (1-x)$.
It vanishes at the endpoints unless the interaction is momentum-independent.)
We therefore write
\begin{equation}
\varphi_\pi^{G_s}(x) = x^{\alpha_-} (1-x)^{\alpha_-}
\bigg[ 1 + \sum_{2,4,\ldots}^{j_s} a_j^\alpha C_j^\alpha(2 x -1) \bigg],
\end{equation}
and minimise $\varepsilon_s = \sum_{m=1,\ldots,m_{\rm max}} |\langle x^m\rangle^{G_s}/\langle x^m\rangle-1|$.  A value of $j_s=2$ ensures mean-\{$|\langle x^m\rangle^{G_{s+2}}/\langle x^m\rangle^{G_{s}}-1||m=1,\ldots,m_{\rm max}\}< 1$\%.
In using Gegenbauer-$\alpha$ polynomials we allow the PDA to differ from $\varphi_\pi^{\rm asy}$ for any finite $\zeta$ and accelerate the procedure's convergence by optimising $\alpha$.  One may project our result onto a $\{C_n^{3/2}\}$-basis, which is that used by other authors, but this incurs costs: requiring far more nonzero coefficients, $\{a^{3/2}_j\}$, and introducing spurious oscillations that are typical of Fourier-like approximations to a simple function.

\begin{figure}[t]
\includegraphics[width=0.85\linewidth]{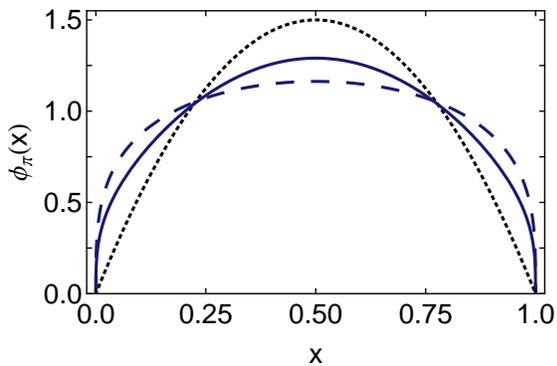}
\caption{
%
Computed distribution amplitude at $\zeta=2\,$GeV.
Curves: solid, DCSB-improved kernel (DB); dashed, rainbow-ladder (RL); and dotted, asymptotic distribution.
\label{fig:phiplot}}
\end{figure}

The dashed curve in Fig.~\ref{fig:phiplot} is our RL result, obtained with $D\omega = (0.87\,{\rm GeV})^3$, $\omega=0.5\,$GeV.  It is described by
\begin{equation}
\label{resphipi2}
\varphi_\pi^{\rm RL}(x) = 1.74 [x (1-x)]^{\alpha_-^{\rm RL}} \, [1 + a_2^{\rm RL} C_2^{\alpha_{\rm RL}}(2 x - 1)]\,,
\end{equation}
with $\alpha_{\rm RL} = 0.79$, $a_2^{\rm RL}=0.0029$.  Projected onto a Gegenbauer-$(\alpha=3/2)$ basis, Eq.\,\eqref{resphipi2} corresponds to $a_2^{(3/2)}=0.23$, \ldots, $a_{14}^{(3/2)}=0.022$, etc.  That $j\geq 14$ is required before $a_{j}^{(3/2)}<0.1\,a_2^{(3/2)}$ highlights the merit of reconstruction via Gegenbauer-$\alpha$ polynomials at any reasonable scale, $\zeta$.  The merit is greater still if, as in lattice-QCD, one only has access to a single nontrivial moment.  In seeking an estimate of $\varphi_\pi(x)$, it is better to fit $\alpha$ than to force $\alpha=3/2$ and infer a value for $a_2^{(3/2)}$.

The solid curve in Fig.\,\ref{fig:phiplot}, described by
\begin{equation}
\label{resphipi2DB}
\varphi_\pi^{\rm DB}(x) = 1.81 [x (1-x)]^{\alpha_-^{\rm DB}} \, [1 + a_2^{\rm DB} C_2^{\alpha_{\rm DB}}(2 x - 1)]\,,
\end{equation}
$\alpha_{\rm DB} = 0.81$, $a_2^{\rm DB}=-0.12$, was obtained using the most sophisticated symmetry-preserving DSE kernels that are currently available \cite{Chang:2011ei}, with $D\omega = (0.55\,{\rm GeV})^3$, $\eta=0.6$.  Projected onto a $\{C_n^{3/2}\}$-basis,
Eq.\,\eqref{resphipi2DB} 
corresponds to $a_2^{(3/2)}=0.15$.  Only for $j\geq 14$ is $a_{j}^{(3/2)}<0.1 \, a_2^{(3/2)}$.

By way of context, we note that a computation using QCD sum rules \cite{Braun:1988qv} produced $\varphi_\pi(x=1/2) = 1.2 \pm 0.3$, which may be compared with:
$\varphi_\pi^{\rm RL}(1/2) = 1.16\,$,
$\varphi_\pi^{\rm DB}(1/2) = 1.29\,$;
and with the value from the asymptotic form, $\varphi_\pi^{\rm asy}(1/2) = 1.5$.  In addition, we find
\begin{equation}
\langle (2x-1)^2 \rangle^{\rm RL} = 0.28\,,\;
\langle (2x-1)^2 \rangle^{\rm DB} = 0.25\,.
\end{equation}
Lattice-QCD \cite{Braun:2006dg} yields a value of $0.27\pm 0.04$ for this moment, whereas it is $0.2$ for the asymptotic distribution.

Numerous qualitatively significant results can be read from Fig.\,\ref{fig:phiplot}.  
The most important being that DCSB is expressed in the PDA through a marked broadening with respect to $\varphi_\pi^{\rm asy}$.  This may be claimed because we have computed the PDA at a low renormalisation scale in the chiral limit, whereat the quark mass function owes entirely to DCSB; and, on the domain $0<p^2<\zeta^2$, the nonperturbative interactions responsible for DCSB produce significant structure in the dressed-quark's self-energy.  The PDA is an integral of the pion's Bethe-Salpeter wave-function, whose pointwise behaviour is rigorously connected with that of the quark self-energy (see Eq.\,\eqref{gtE} and kindred Goldberger-Treiman relations \cite{Maris:1997hd}).  Hence, the structure of the pion's distribution amplitude at the hadronic scale is a pure expression of DCSB.  As the scale is removed to extremely large values, phase space growth diminishes the impact of nonperturbative DCSB interactions, so that the PDA relaxes to its asymptotic form.

Signficant, too, is the pointwise difference between the DB and RL results.  It is readily understood, bearing in mind that low-$m$ moments are most sensitive to $\varphi_\pi(x)$ in the neighbourhood of $x=1/2$, whereas high-$m$ moments are sensitive to its endpoint behaviour.
RL-kernels ignore DCSB in the quark-gluon vertex.  Therefore, to describe a given body of phenomena, they must shift all DCSB-strength into the infrared behaviour of the quark propagator, whilst nevertheless maintaining perturbative behaviour for $p^2>\zeta^2$.  This requires $B(p^2)$ to be large at $p^2=0$ but drop quickly, behaviour which influences $\varphi_\pi(x)$ via Eq.\,\eqref{gtE}.  The concentration of strength at $p^2\simeq 0$ forces large values for the small-$m$ moments, which translates into a broad distribution.
In contrast, the DB-kernel builds DCSB into the quark-gluon vertex and its impact is therefore shared between more elements of a calculation.  Hence a smaller value of $B(p^2=0)$ is capable of describing the same body of phenomena; and this self-energy need fall less rapidly in order to reach the common asymptotic limit.  (Using Eqs.\,\eqref{SgeneralN}, these remarks become evident in Fig.\,\ref{fig:Splot}.)  It follows that the low-$m$ moments are smaller and the distribution is narrower.
Both PDAs have the same large-$x$ behaviour because the RL and DB kernels agree at ultraviolet momenta.

Notably, one should not expect to obtain agreement with data for a given process by using our computed form of $\varphi_\pi(x)$ in the relevant lowest-order (in coupling), leading-twist formula.  This is illustrated well via the $\gamma^\ast \gamma \to \pi^0$ transition form factor, $G_{\gamma \pi \gamma^\ast}$.  The dashed curve in Fig.\,\ref{fig:phiplot} was obtained using a RL DSE kernel in that class which reproduces all uncontroversial data on this process \cite{Roberts:2010rnS}.  However, when employed in the asymptotic formula \cite{Lepage:1980fj}, the result for $Q^2 G_{\gamma \pi \gamma^\ast}$ is too large by roughly a factor of two.  Plainly, subleading contributions are important, at least for $Q^2\lesssim 10\,$GeV$^2$ and probably on a larger domain, as also observed elsewhere \cite{Brodsky:2011yv,Bakulev:2012nh}.

Our PDA computations unify a diverse range of phenomena.  The rainbow-ladder result, e.g., connects directly with \emph{ab initio} predictions for: $\pi \pi$ scattering, and pion electromagnetic elastic and transition form factors \cite{Maris:2003vk}; and nucleon and $\Delta$ properties \cite{Eichmann:2011ej}.  And, although use of DCSB-improved kernels is just beginning, our related prediction for the PDA links immediately with analyses showing that DCSB is, e.g., responsible for both a large dressed-quark anomalous magnetic moment \cite{Chang:2010hb} and the splitting between parity partners in the spectrum \cite{Chang:2011ei,Chen:2012qrS}.

The pion's PDA is the closest thing in QCD to a quantum mechanical wave function for the pion.  Its hardness at an hadronic scale is a direct expression of DCSB.

\smallskip

\noindent\textbf{Acknowledgments}.
Work supported by:
For\-schungs\-zentrum J\"ulich GmbH;
University of Adelaide and Australian Research Council through grant no.~FL0992247;
%
Department of Energy, Office of Nuclear Physics, contract no.~DE-AC02-06CH11357;
and 
National Science Foundation, grant no.\
NSF-PHY-0903991.

\smallskip

\begin{table}[t]
\caption{Representation parameters. \emph{Upper panel}: Eq.\,\protect\eqref{Spfit} -- the pair $(x,y)$ represents the complex number $x+ i y$.  \emph{Lower panel}: Eqs.\,\protect\eqref{Fifit}--\protect\eqref{Gufit}.  (Dimensioned quantities in GeV).
\label{Table:parameters}
}
\begin{center}
%

\begin{tabular*}
{\hsize}
{
l|@{\extracolsep{0ptplus1fil}}
c@{\extracolsep{0ptplus1fil}}
c@{\extracolsep{0ptplus1fil}}
c@{\extracolsep{0ptplus1fil}}
c@{\extracolsep{0ptplus1fil}}
c@{\extracolsep{0ptplus1fil}}}\hline
 RL & $z_1$ & $m_1$  & $z_s$ & $m_2$ \\

    & $(0.44,0.014)$ & $(0.54,0.23)$ & $(0.19,0)$ & $(-1.21,-0.65)$ \\
 DB & $z_1$ & $m_1$  & $z_s$ & $m_2$ \\
    & $(0.44,0.28)$ & $(0.46,0.18)$ & $(0.12,0)$ & $(-1.31,-0.75)$ \\\hline
\end{tabular*}

\begin{tabular*}
{\hsize}
{
l@{\extracolsep{0ptplus1fil}}
l@{\extracolsep{0ptplus1fil}}
c@{\extracolsep{0ptplus1fil}}
c@{\extracolsep{0ptplus1fil}}
c@{\extracolsep{0ptplus1fil}}
c@{\extracolsep{0ptplus1fil}}
c@{\extracolsep{0ptplus1fil}}
c@{\extracolsep{0ptplus1fil}}
c@{\extracolsep{0ptplus1fil}}
c@{\extracolsep{0ptplus1fil}}}\hline
    & & $c^{\rm i}$ & $c^{u}$ & $\phantom{-}\nu^{\rm i}$ & $\nu^{\rm u}$ & $a$\phantom{00} & $\Lambda^{\rm i}$ & $\Lambda^{\rm u}$\\\hline
RL: & E & $1 - c^{u}_E$ & $0.03$ & $-0.71$ & 1.08
    & 2.75\phantom{$/[\Lambda^{\rm i}_G]^3$} & 1.32 & 1.0\\
    & F & 0.51 & $c^{\rm u}_E/10$ & $\phantom{-}0.96$ & 0.0
    & 2.78$/\Lambda^{\rm i}_{F}$\phantom{00} & 1.09 & 1.0 \\
& G & $0.18$ & 2$\,c^{\rm u}_F$ & $\phantom{-}\nu^{\rm i}_F$ & 0.0 & 5.73$/[\Lambda^{\rm i}_G]^3$ & 0.94 & 1.0 \\\hline
DB: & E & $1 - c^{u}_E$ & $0.08$ & $-0.70$ & 1.08
    & 3.0\phantom{$/[\Lambda^{\rm i}_G]^3$} & 1.41 & 1.0\\
    & F & \phantom{-}0.55 & $c^{\rm u}_E/10$ & $\phantom{-}0.40$ & 0.0
    & 3.0$/\Lambda^{\rm i}_{F}$\phantom{00} & 1.13 & 1.0 \\
& G & $-0.094$ & 2$\,c^{\rm u}_F$ & $\phantom{-}\nu^{\rm i}_F$ & 0.0 & 1.0$/[\Lambda^{\rm i}_G]^3$ & 0.79 & 1.0 \\\hline
\end{tabular*}
\end{center}

\vspace*{-4ex}

\end{table}
\noindent\textbf{Appendix}.
Here we describe the interpolations used in our evaluation of the moments in Eq.\,\eqref{phimom}.  The dressed-quark propagator is represented as \cite{Bhagwat:2002tx}
\begin{equation}
S(p) = \sum_{j=1}^{j_m}\bigg[ \frac{z_j}{i \gamma\cdot p + m_j}+\frac{z_j^\ast}{i \gamma \cdot p + m_j^\ast}\bigg], \label{Spfit}
\end{equation}
with $\Im m_j \neq 0$ $\forall j$, so that $\sigma_{V,S}$ are meromorphic functions with no poles on the real $p^2$-axis, a feature consistent with confinement \cite{Bashir:2012fs}.  We find that $j_m=2$ is adequate.

With relative momentum defined via $\eta=1/2$, we represent the scalar functions in Eq.\,\eqref{genGpi} $({\cal F}=E,F,G)$ by
{\allowdisplaybreaks
\begin{eqnarray}
{\cal F}(k;P) &=& {\cal F}^{\rm i}(k;P) + {\cal F}^{\rm u}(k;P) \,, \\
\nonumber {\cal F}^{\rm i}(k;P) & = & c_{\cal F}^{\rm i}\int_{-1}^1 \! dz \, \rho_{\nu^{\rm i}_{\cal F}}(z) \bigg[
a_{\cal F} \hat\Delta_{\Lambda^{\rm i}_{{\cal F}}}^4(k_z^2) \\
&& \rule{7em}{0ex}
+ a^-_{\cal F} \hat\Delta_{\Lambda^{\rm i}_{\cal F}}^5(k_z^2)
\bigg], \label{Fifit}\\
E^{\rm u}(k;P) & = & c_{E}^{\rm u} \int_{-1}^1 \! dz \, \rho_{\nu^{\rm u}_E}(z)\,
 \hat \Delta_{\Lambda^{\rm u}_{E}}(k_z^2)\,,\\
F^{\rm u}(k;P) & = & c_{F}^{\rm u} \int_{-1}^1 \! dz \, \rho_{\nu^{\rm u}_F}(z)\,
 \Lambda_F^{\rm u} k^2 \Delta_{\Lambda^{\rm u}_{F}}^2(k_z^2)\,,\\
G^{\rm u}(k;P) & = & c_{G}^{\rm u} \int_{-1}^1 \! dz \, \rho_{\nu^{\rm u}_G}(z)\,
 \Lambda_G^{\rm u}\Delta_{\Lambda^{\rm u}_{G}}^2(k_z^2)\,, \label{Gufit}
\end{eqnarray}}
\hspace*{-0.5\parindent}with $\hat \Delta_\Lambda(s) = \Lambda^2 \Delta_\Lambda(s)$, $k_z^2=k^2+z k\cdot P$, $a^-_E = 1 - a_E$, $a^-_F = 1/\Lambda_F^{\rm i} - a_F$, $a^-_G = 1/[\Lambda_G^{\rm i}]^3 - a_G$.  $H(k;P)$ is small, has little impact, and is thus neglected.

Values of the interpolation parameters that fit our numerical results are presented in Tables~\ref{Table:parameters}.  Those for the Bethe-Salpeter amplitudes were obtained through a least-squares fit to the Chebyshev moments
\begin{equation}
{\cal F}_n(k^2) = \frac{2}{\pi}\int_{-1}^{1}\!dx\, \sqrt{1-x^2} {\cal F}(k;P) U_n(x)\,,
\end{equation}
with $n=0,2$, where $U_n(x)$ is an order-$n$ Chebyshev polynomial of the second kind. Owing to $O(4)$ invariance, we may define $x = \hat k\cdot P/ip$, with $\hat k^2=1$ and $P=(0,0,p,i p)$.

The strength of the interaction detailed in Ref.\,\cite{Qin:2011ddS} is specified by a product: $D\omega = m_G^3$.  With $m_G$ fixed, results for properties of ground-state vector and flavour-nonsinglet pseudoscalar mesons are independent of the value of $\omega \in [0.4,0.6]\,$GeV.  We use $\omega =0.5\,$GeV.
With the RL kernel, $f_\pi=0.092\,$GeV is obtained with $m_G^{\rm RL}(2\,GeV)=0.87\,$GeV and $m_G^{\rm RL}(19\,GeV)=0.80\,$GeV, whilst with the DB kernel it is obtained with $m_G^{\rm DB}(2\,GeV)= m_G^{\rm DB}(19\,GeV)=0.55\,$GeV.
Plainly, multiplicative renormalisability is better preserved with the DB kernels.
In Eq.\,(10) of Ref.\,\cite{Chang:2011ei}, the strength of the dressed-quark anomalous chromomagnetic moment was described by a value $\eta=0.65$.  To improve numerical stability in the interpolations described herein, we changed to $\eta=0.6$.  This increases the computed value of the $a_1$-$\rho$ mass-splitting by less-than $15$\%.


\end{document}